\documentclass[10pt, conference, compsocconf]{IEEEtran}

\usepackage{float}
\usepackage{wrapfig}
\usepackage{etoolbox}
\usepackage[usenames,dvipsnames,svgnames,x11names,table]{xcolor}

\usepackage[T1]{fontenc}
\usepackage{microtype}
\usepackage{listings}
\usepackage{calrsfs}
\usepackage{siunitx} % Make units such as degrees
\usepackage{amsmath} % Be careful if clashes
\usepackage{amssymb}
\usepackage{mathrsfs}
\usepackage{latexsym}
\usepackage{mathptmx}
\usepackage{graphicx}
\usepackage{tabularx}
\usepackage{xkeyval}
\usepackage{adjustbox}
\usepackage{caption}
\usepackage{subcaption}

\usepackage{tikz}

\usepackage[backend=biber,style=ieee,firstinits=true,maxbibnames=99,uniquename=false]{biblatex}
\usepackage{varioref} % Declare before hyperref, for referencing objects in different pages (then put the page number), or same (then do nothing)
\usepackage{hyperref}
\hypersetup{
    bookmarks=true,            % show bookmarks bar?
    unicode=true,              % non-Latin characters in Acrobat’s bookmarks
    pdfnewwindow=false,        % links in new window
    colorlinks=true,           % false: boxed links; true: colored links
    linkcolor=DeepSkyBlue1,    % color of internal links (change box color with linkbordercolor)
    citecolor=SlateBlue2,      % color of links to bibliography
    filecolor=DarkOliveGreen2, % color of file links
    urlcolor=DodgerBlue3       % color of external links
}
\usepackage[nameinlink]{cleveref}

\begin{document}

\title{Agent based network modelling of COVID-19 disease dynamics and vaccination uptake in a New South Wales Country Township}
\author{\IEEEauthorblockN{Shing Hin (John) Yeung\IEEEauthorrefmark{1}} \IEEEauthorblockA{Faculty of Engineering, University of Sydney
		\\\texttt{syeu2887@uni.sydney.edu.au}} \and \IEEEauthorblockN{Mahendra Piraveenan\IEEEauthorrefmark{2}} 
	\IEEEauthorblockA{Faculty of Engineering, University of Sydney
		\\\texttt{mahendrarajah.piraveenan@sydney.edu.au}
	    \\ \href{https://orcid.org/0000-0001-6550-5358}{ORCID}}
	} 
\maketitle

\begin{abstract}

We employ an agent-based contact network  model to study the relationship between vaccine uptake  and disease dynamics in a hypothetical country town from New South Wales, Australia, undergoing a COVID-19 epidemic, over a period of three years.  We model the contact network in this hypothetical township of  $N\;=\;10000$ people as a scale-free network, and simulate the spread of COVID-19 and vaccination program using disease and vaccination uptake parameters typically observed in such a NSW town. We simulate the spread of the ancestral variant of COVID-19 in this town, and study the disease dynamics while the town maintains limited but non-negligible contact with the rest of the country which is assumed to be undergoing a severe COVID-19 epidemic.  We also simulate a maximum three doses of  Pfizer Comirnaty vaccine being administered in this town, with limited vaccine supply at first which gradually increases, and analyse how the vaccination uptake affects the disease dynamics in this town, which is captured using an extended compartmental model with  epidemic parameters typical for a COVID-19 epidemic in Australia. Our results show that, in such a township, three vaccination doses are sufficient to contain but not eradicate COVID-19, and the disease essentially becomes endemic. We also show that the average degree of infected nodes (the average number of contacts for infected people) predicts the proportion of infected people. Therefore, if the hubs (people with a relatively high number of contacts) are disproportionately infected, this indicates an oncoming peak of the infection, though the lag time thereof depends on the maximum number of vaccines administered to the populace. Overall, our analysis provides interesting insights in understanding the interplay between network topology, vaccination levels, and COVID-19 disease dynamics in a typical remote NSW country town.

\end{abstract}

\section[Introduction]{Introduction}

Since first being reported in  December 2019, Severe Acute Respiratory Syndrome Coronavirus 2 (SARS-CoV-2) has spread globally, and caused the Coronavirus Disease 2019 (COVID-19) pandemic. As of April 2023,  this pandemic has been directly responsible for $763$ million cases and $6.3$ million deaths \cite{WorldData2023}.  The World Health Organisation (WHO) has  advocated a holistic approach in public policies to control the pandemic \cite{WHO2023}.  A cornerstone of this holistic approach is ensuring high levels of vaccination in communities against COVID-19.

In this work, we study the effect of multiple doses of vaccination in a small community. Specifically, we are interested in finding out how many doses of vaccine are needed in  a small community, for a particular set of typical disease parameters, before the disease can be eradicated from or contained in that community. We are also interested in analysing how vaccination of highly connected people (`hubs' in the underlying contact network) can influence vaccination threshold to achieve herd immunity. We consider a typical New South Wales township with ten thousand inhabitants, who have a limited level of contact with the rest of  the world, throughout the epidemic,  who receive their first supply of vaccines a certain time after the epidemic begins in the township, and have limitations on vaccine supply at the beginning of the vaccination program. We assume that the epidemic is seeded in the town through external contact.  We model the contact network as a scale-free network and we simulate the spread of COVID-19 using disease parameters typical of the ancestral variant of SARS-CoV-2.  We then simulate a realistic vaccination program and study the effects of sequential vaccination on the population in terms of disease dynamics and disease eradication.

We find that the ability  of the vaccination program to  eradicate COVID19 from the community depends on both the number of vaccines administered, as well as the average period of natural immunity people acquire from being infected by SARS-CoV-2. When the period of natural immunity is assumed to be 180 days or higher, it is possible to significantly eradicate (bring  the proportion of infected people down to 10\% or lower) COVID19 by administering three doses of the vaccine. However, when the natural immunity time is much lower, the disease becomes endemic even when subsequent doses of vaccine are administered, assuming these vaccines are also administered at intervals of 180 days, after the second vaccine.  That is, it is possible to prevent the disease becoming endemic only if the duration between boosters is significantly smaller than the period of natural immunity from disease. This makes sense since vaccination uptake is not assumed to be 100 \%, and if those  who have  not taken the vaccine or boosters become susceptible at the same time as those who have taken vaccines and their vaccine immunity is wearing off, then the disease will spread again and cannot be eradicated. Furthermore, considering the role of the highly connected people (hubs in the underlying contact network),  we find that there is strong correlation between the average degree of the infected nodes (people), and the number of infected people. That is, when nodes which have higher degree (hubs) are infected, this results in the infection peaking. We further verified this by running Granger causality tests and calculating lag times, which showed that when there  was no vaccination the lag time (between these two events) was longer. When vaccines were administered, the lag times become shorter. In short, in such a small isolated community, the average degree of the infected people can predict peaks in infection. Overall, our results demonstrate the interdependencies between contact network topology, vaccination, and disease dynamics, in a small isolated town which has small but non-negligible traffic volume with the outside world.

\section[Background]{Background}

\subsection{Basic Reproduction Number}

The basic reproduction number, $R_0$, for a disease is the average number of people who contract the disease from a single person in a population which is entirely susceptible \cite{Chang2020}.   In this study, we calculate $R_0$ as the ratio of the transmission  rate and recovery rate $R_0\;\equiv\;\frac{\beta}{\delta}$.  The estimated $R_0$ for the COVID-19 caused by the ancestral variant of SARS-CoV-2 ranged between $2.5$-$3.0$ \cite{Liu2020}.  Meanwhile,  the delta variant has an $R_0$ of $3.2$-$6.0$ \cite{Liu2021}. Similarly the omicron has an $R_0$ of $1.9$-$9.5$ \cite{Manathunga2023} depending on subvariants. 

\subsection{Vaccination against COVID-19}

The COVID-19 vaccines arrived on the market in approximately December 2020, well within the expected duration of development \cite{Piraveenan2021}.  The COVID-19 vaccines range from recombinant viral-vectored vaccines to mRNA-based vaccines, including (i) the Pfizer-BioNTech BNT162  mRNA vaccine, (ii) Spikevax from Moderna, which is designed to induce antibodies against a portion of the coronavirus spike protein \cite{Jeyanathan2020}, (iii) Nuvaxovid, which contains Matrix-M adjuvant to induce antibodies similar to Spikevax \cite{DH2023a} (iv) the ChAdOx1  viral vector vaccine  by the University of Oxford and AstraZeneca (v) an inactivated whole virus vaccine by Sinovac. The use of  these vaccines  provides a safe pathway  to contain the  COVID-19 disease before a comprehensive treatment is developed, and prevents a sufficient proportion of the population being infected to transmit the virus broadly, especially to the vulnerable  sections of the community. 

At the time of writing, the COVID-19 pandemic has not ended, and the  WHO recommends that communities continue to take booster vaccines to keep up immunity \cite{WHO2023}.   In such a context, it becomes important to see the effect of continued vaccination and how it can help to contain and eventually eradicate COVID-19 from communities. Since such a task is manifestly easier in isolated communities, in this work we consider an imaginary remote township in country NSW, Australia, which has limited (but not negligible) connections with the rest of the country and world. However, it should be emphasised that the model could be easily  calibrated to represent other remote communities.

\subsection{Related work}

Indeed, a vast body of work exists that deals with computational modelling of COVID-19. Quite often, agent-based approaches and synthetic populations \cite{adiga2015generating} are used in such modelling. For example, Kerr et al. \cite{Ker2021} and Wolfram \cite{wolfram2020} use agent-based models to analyse COVID-19 disease dynamics and prescribe intervention strategies. The contact network aspect of COVID-19 spread also has been dealt with in a number of ways, such as by considering proximity metrics \cite{Mehrab2022} or Heterogeneous adaptive behavioral responses  \cite{Espinoza2022}.  The primary focus of this work, however, is to focus on a fairly isolated country town which has limited (but not negligible) contact with the outside world, and focus on modelling disease dynamics in such a town using agent-based modelling and contact network modelling.

\section{Method}

In this work, we create an agent-based network model of a small  New South Wales (NSW) township which has a population of $N\;=\;10000$. The epidemic is seeded with four initial infections, which was incidentally how COVID-19 was seeded in Australia in 2020 \cite{NSWHealth2020}.  The seeds are chosen randomly.  The contact network was modelled as a scale-free network. The infection dynamics is  simulated for $T\;=\;1080$ days, that is approximately three years.  Note that this period is chosen to reflect the time period between 2020 -  2022, the three years during which Australian country towns were badly affected by COVID-19 \cite{stobart2022australia}.  The simulation is conducted from a self-authored Python package (\url{https://github.com/lt-shy-john/covid19-vaccine-game-theory}).

\subsection{Compartment Model}
The compartment model we employ is based on the work of  Abou-Ismail et al\cite{Abou-Ismail2020}.  Specifically, the model contains  Susceptible ($S$), Exposed ($E$),  Quarantined  ($U$), Recovered  ($R$), and Vaccinated ($V$) compartments.   Exposed people are assumed to be infectious, but once quarantined, they will not pass on the pathogen to others.  The Infected compartment also can be considered, which is  the sum of Exposed and Quarantined compartments (i.e. $I\;=\;E\;+\;U$). The vaccinated compartment can be divided into people who got  immunity from one vaccine ($V_1$), two vaccines ($V_2$), three vaccines  ($V_3$) etc. In this study, we consider a maximum of three vaccinations.  Vaccine efficacy is  explicitly modelled, as described later, and people are considered to be in the `Vaccinated' compartments only if they received immunity from the corresponding vaccine (not simply by receiving the vaccine).  Susceptible people become exposed with a transmission rate of $\beta$,  and exposed people become Quarantined with predefined rates which are dependent on either testing rate  $\lambda$  or the incubation period of the disease ($\frac{1}{\tau}$). The recovery rate from being infected is $\delta$, and the rate of recovered people becoming susceptible after recovery  is  $\gamma$ (that is, natural immunity conferred by the disease lasts for a period of $\frac{1}{\gamma}$ on average). The vaccination rate is $\alpha_1$ for  people getting one vaccine ($V_1$),   $\alpha_2$ for people moving to two vaccines ($V_2$),   $\alpha_3$ for people moving to three vaccines  ($V_3$) etc. In this study, we consider a maximum of three vaccinations.  Similarly,   $\phi_i$ is the rate in which people who have got the  $i^{th}$ vaccination lose immunity conferred by that vaccine, and become  Susceptible again. This  model is shown in \cref{fig:SIRV} below. 

In this model, a person in the exposed compartment may either (i) discover they are infected via a COVID-19 test, subject to a testing rate $\lambda$ which denotes the testing probability of a person before they become symptomatic, (ii) discover they are infected via becoming symptomatic, after an incubation period of ($\frac{1}{\tau}$).   Either mechanism will see them enter the quarantined compartment ($U$), therefore the rate in which people move from Exposed compartment ($E$) to Quarantined compartment  ($U$) is  $\lambda + \tau$.  We assume that  at this point, the person  who is infected will  no longer be able to transmit to others (e.g.  will be hospitalised or  will be in self-isolation). For this study, we assume  $R_0\;=\;2.6$.

In this study, we assume that  the second vaccine is administered exactly 28 days after the first vaccine to those who are willing, and the third vaccine is administered exactly 180 days after the second. In reality, these periods will vary, but the above periods are recommended \cite{DH2023b} and variations thereof are assumed to even out on average. The initial conditions and parameter values  assumed in this study are  given in Table \cref{table:para}. It is important to note that  we have not conducted a sensitivity analysis, and thus while these parameters are chosen to realistically match the disease conditions in a typical Australian outback town, as mentioned above, a different set of parameter values could result in considerably different disease dynamics compared to that presented below.

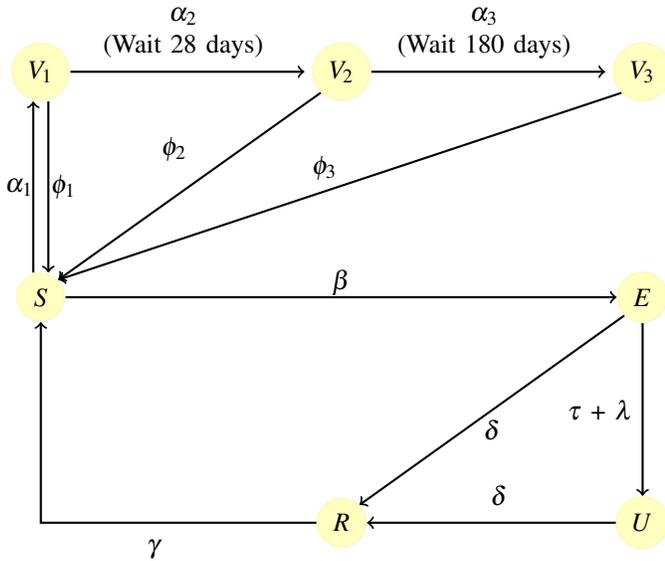
\begin{figure}[h!]
	\centering
	\vspace{-2mm} 
	\begin{tikzpicture}[
		block/.style={%
			circle,minimum size=5mm,thick,draw=Yellow!85!Grey!25!,fill=Yellow!25!
		}]
		\node (S) [block] at (-3,0) {$S$};
		\node (V1) [block] at (-3,3) {$V_1$};
		\node (V2) [block] at (1,3) {$V_2$};
		\node (V3) [block] at (5,3) {$V_3$};
		\node (I) [block] at (5,-3) {$U$};
		\node (E) [block] at (5,0) {$E$};
		\node (R) [block] at (1,-3) {$R$};
		\draw[->,thick] ([xshift=-0.1cm]S.north) {} -- node[right,xshift=-0.5cm] {$\alpha_1$} ([xshift=-0.1cm]V1.south);
		\draw[->,thick] (V1.east) {} |- node[above,xshift=1.5cm] {\begin{tabular}{c} $\alpha_2$ \\ (Wait 28 days) \end{tabular}} ([xshift=-0.1cm]V2.west);
		\draw[->,thick] (V2.east) {} |- node[above,xshift=1.5cm] {\begin{tabular}{c} $\alpha_3$ \\ (Wait 180 days) \end{tabular}} ([xshift=-0.1cm]V3.west);
		\draw[->,thick] ([xshift=0.1cm]V1.south) {} -- node[left,xshift=0.5cm] {$\phi_1$} ([xshift=0.1cm]S.north);
		\draw[->,thick] (V2.south west) {} -- node[right,xshift=-0.5cm,yshift=0.5cm] {$\phi_2$} (S.north east);
		\draw[->,thick] (V3.south west) {} -- node[right,xshift=-0.5cm,yshift=0.25cm] {$\phi_3$} (S.north east);
		\draw[->,thick] (S.east) {} -- node[below,yshift=0.5cm] {$\beta$} (E.west);
		\draw[->,thick] (E.south) -- node[left] {$\tau$  + $\lambda$} (I.north);
		\draw[->,thick] (I.west) {} -- node[above,xshift=0.1cm,yshift=0.1cm] {$\delta$} (R.east);
		\draw[->,thick] (E.south west) {} -- node[below] {$\delta$} (R.north east);
		\draw[->,thick] (R.west) {} -| node[below,xshift=1.5cm,yshift=-0.1cm] {$\gamma$} (S.south);
	\end{tikzpicture}
	\caption{The SIRV  Compartmental Model for COVID-19 disease dynamics (similar to the one in Abou-Ismail et al \cite{Abou-Ismail2020}), consisting Susceptible ($S$), Exposed ($E$),  Quarantined  ($U$), Recovered  ($R$), and Vaccinated ($V$) compartments. Three doses of vaccination are assumed, the second and the third being administered 28 and 208 days after the first.  The transmission rate is $\beta$, recovery rate is $\delta$, the rate of becoming susceptible after disease is  $\gamma$, the vaccination rates are  $\alpha_i$ for  $i^{th}$ vaccination, calculated as a proportion of people who took the previous vaccine, and the rate of vaccination immunity wearing off from $i^{th}$ vaccination is $\phi_i$.}
	\label{fig:SIRV}
	\vspace{-3mm} 
\end{figure}

\begin{table}[h!]
	\centering
	\begin{tabularx}{0.45\textwidth}{ | X | c | c | }
		\hline
		\emph{Parameter} & \emph{Symbol} & \emph{Value} \\ 
		\hline
		Population & $N$ & 10000 people\\  
		\hline
		Duration & $T$ & 1080 days \\  
		\hline
		Initial infection &  & 4 people\\ 
		\hline
		\multicolumn{3}{|c|}{SIRV Model} \\
		\hline
		Vaccine adoption rate & $\alpha (\alpha_1, \alpha_2, \alpha_3)$ & 0.9 \\  
		\hline
		Transmission rate & $\beta$ & 0.14 \\
		\hline
		Recovery rate & $\delta$ & $0.05$ \\
		\hline
		Vaccine wear-off  rates & $\phi  (\phi_1, \phi_2, \phi_3)$ & $0.0056  (1/180) $ \\
		\hline
		Testing rate & $\lambda$ & $0.01$/ day \\
		\hline
		Incubation period & $\frac{1}{\tau}$ & $14$ days \\
		\hline
		Natural immunity period & $\frac{1}{\gamma}$ & $180$ days \\
		\hline
		\multicolumn{3}{|c|}{Network Model} \\
		\hline
		Average degree & $\left\langle k\right\rangle$ & $ 5.00$ \\
		\hline
		Scale-free exponent & $\omega$ & $2.05$ \\
		\hline
		\multicolumn{3}{|c|}{Out-of-town Travel} \\
		\hline
		Departure rate & $\phi_d$ & $0.00012$ \\
		\hline
		Return rate & $\phi_r$ & $0.0001$ \\
		\hline
	\end{tabularx}
\caption{Simulation parameters. }
\label{table:para}
\vspace{-3mm} 
\end{table}

\subsection{Social/ Contact Network Model}

We adapt scale-free networks \cite{BarabasiPosfai2016}  to model contact networks. Scale-free networks are ubiquitous in real world, and for this reason often used as model networks in various contexts (e.g. \cite{BarabasiPosfai2016}).  Scale-free networks have been shown to be good models of human contact networks (e.g. \cite{Liu2021}). In a scale-free network, the degree distribution follows a power law, and the probability of a node to have a degree  of $k$ is given by $p_k = Ak^{-\omega}$ where $A$ is a constant and $\omega$ is the power law exponent (also referred to as scale-free exponent) \cite{BarabasiPosfai2016}.  A higher value  of $\omega$ results in a degree distribution with a steeper slope, while  a lower value of $\omega$ results in a  flatter degree distribution.

We use the Barabasi-Albert (BA) growth model \cite{BarabasiPosfai2016} to grow scale-free networks. The scale-free network used in this simulation to model the contact network have $N\;=\;10000$ nodes, an average degree of $\left\langle k\right\rangle\;=\;5.00$, and a scale-free exponent of $\omega\;=\;2.05$, which is calculated with a fitness of 87.8\%.

\subsection{Outside travel}

We assume that a small number of  people travel out of the town, or return to it, at a low rate. The probability of a person  travelling out of the town on a given day is modelled as $\phi_d = 0.00012$, and the probability of a person outside the town returning to it on a given day is modelled as $\phi_r = 0.0001$.  We assume that the travel rate is similar to Australians in general leaving or returning to Australia during the pandemic, and thus the travel and return probabilities are calculated from the number of Australians who departed from  and retuned to the country between Feb 2022 and April 2023 \cite{ABS2023a} converted to a daily rate.   

Of course, a person who travels outside the town may contract COVID-19 while outside and bring it into the town, and this is the primary method whereby the infection is further seeded in the town, and the primary reason for the possibility of the disease becoming endemic in the town. Therefore, we assume that people who travel outside the town go to and stay in highly dense population centres, and these population centres are assumed to be well-mixed, rather than having a contact network. We assume that a person who travels outside goes to a population centre with a population of $N_o = 4,000,000$ (population of a typical Australian state capital)  where COVID-19 spreads in a well-mixed population with  a transmission rate ($\beta_o\;=\;0.14$) and recovery rate  ($\delta_o\;=\;0.05$). Thus, a person from the township has a strong chance of being infected during their stay outside the township, and if they do not recover before they return, then they could bring the infection into the town, and will join the `Exposed' ($E$) compartment upon their return. Of course, a traveller with disease-induced or vaccination-induced immunity which has not worn off would not contract COVID-19.

\subsection{Vaccine Supply}

We assume that the town has no vaccines in store when the disease was seeded, and the first vaccine supplies arrive 60 days after the first cases were seeded.  We model the gradual ramping up of vaccine supply. Thus in our model, the town receives $n_{v1} =  10$ vaccines  on the $60$\textsuperscript{th} day, and supply doubles every day thereafter until is capped at  $n_{v5} =  100$ doses on the $65$\textsuperscript{th} day, and  will stay at that rate of supply thereafter. This limits the number of people who can take the vaccine when it first becomes available. The  second dose is administered at least $28$ days after the first dose for a given person (who is willing to take the second dose), and the third dose is administered $180$ days after (to a willing person). Since vaccine supply is steady at the maximum rate by the time the second and third doses are administered, no willing person need to wait for them. The type of vaccine was assumed to be Pfizer Comirnaty (BNT162b2)  mRNA vaccine, and vaccine efficacy rate is assumed to be $\mu_1 =  0.92$ for the first dose,  $\mu_2  = 0.86$ for the second dose, and  $\mu_3 = 0.96$ for the third dose, which are based on empirical studies \cite{Lancet2021, UKHSA2022}.  The uptake rate is set  as $\alpha_1\;=\alpha_2\;=\alpha_3\;=0.9$ which is similar to the proportion of vaccine takers in Australia \cite{WorldData2023, dodd2021willingness}.  Note well  that this vaccine adoption rate is derived and calibrated from empirical  data and is quite realistic \cite{BORRIELLO2021473, dodd2021willingness}. The vaccination hesitancy is assumed to affect people who are about to take each dose, and a person who has taken the $i^{th}$ dose is assumed to be just as susceptible to vaccine hesitancy as a person who has not taken any vaccination. Thus, while 90\% of the population will take the first vaccine, only 81\% will take the second, and 72.9\% will take the third etc.

\subsection{Simulation Goals}

The goals of the simulation experiments are to (i) recognise whether the COVID-19 could be eradicated from a small township which is largely isolated but yet has limited contact with the rest of the country, or whether the disease will become endemic (ii) if eradication is possible. how many doses of vaccination are necessary to achieve it (iii) if the disease becomes endemic, what is the period of the endemic circle (iv) if targeted vaccination of the hubs or higher-degreed nodes (people who have a high number of contacts) can help in containing the disease. In this initial study we focus on simulating the model with a realistic set of parameters, rather than understanding the effects of varying each of these parameters individually. Therefore we have largely used fixed values  which are realistic for each parameter.

%The primary variable during experiments was the number of vaccinations. 

\section{Results and Discussion}

\subsection{Endemic Cycles}

\begin{figure*}[h!]
	\centering
	\begin{subfigure}{0.28\textwidth}
		\includegraphics[width=\linewidth]{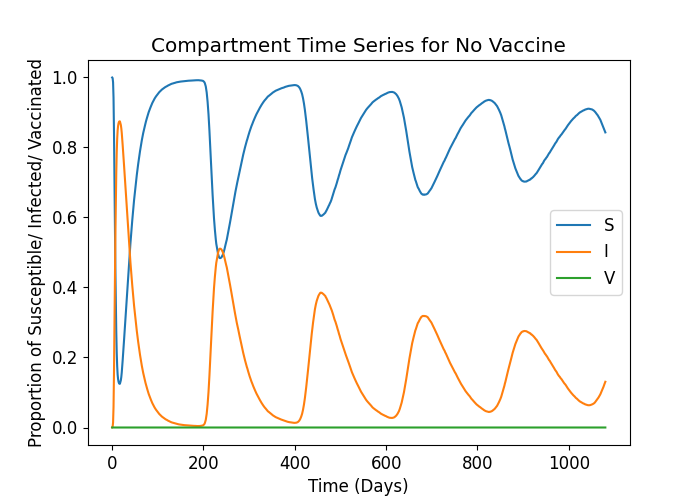}
		\caption{No vaccine provided}
		\label{fig:no_dose_i}
	\end{subfigure}
	~
	\centering
	\begin{subfigure}{0.28\textwidth}
		\centering
		\includegraphics[width=\linewidth]{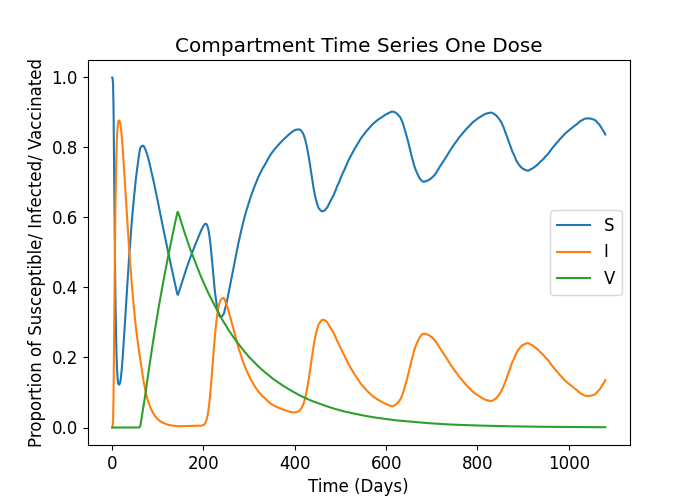}
		\caption{One dose}
		\label{fig:one_dose_i}
	\end{subfigure}
	
	\centering
	\begin{subfigure}{0.28\textwidth}
		\centering
		\includegraphics[width=\linewidth]{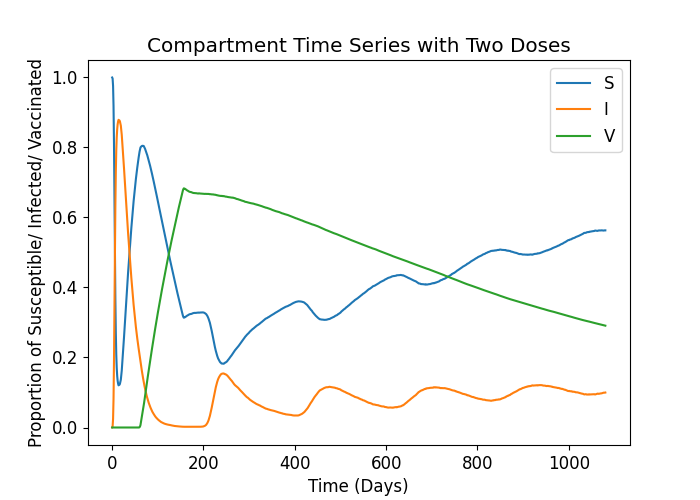}
		\caption{Two doses}
		\label{fig:two_dose_i}
	\end{subfigure}
	~
	\centering
	\begin{subfigure}{0.28\textwidth}
		\centering
		\includegraphics[width=\linewidth]{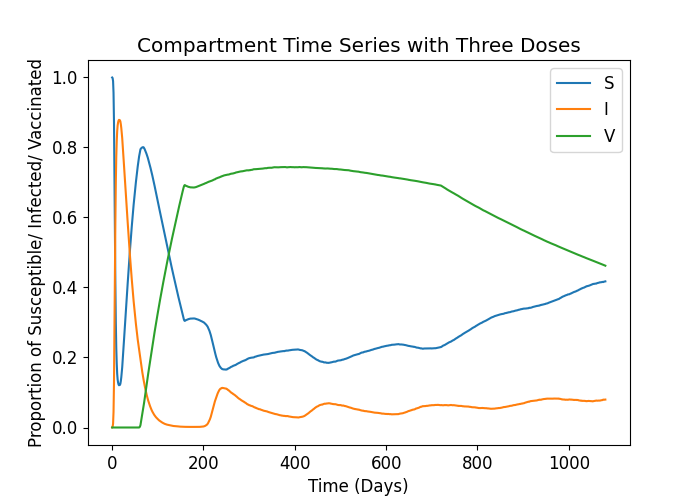}
		\caption{Three doses}
		\label{fig:three_dose_h3_i}
	\end{subfigure}
	\caption{ Infectious disease dynamics of COVID-19 in the NSW township of 10,000 people when a) zero  b) one c) two  d) three vaccines are administered. Second and third vaccines are administered 28 and 208 days after the first, respectively.  The proportion of Susceptible ($S$), Infected ($I =  E + U$) and Vaccinated ($V$) people in the population are shown. Endemic cycles are observed in no vaccine or single vaccine scenarios, whereas the disease is largely contained when three vaccines are administered, even with limited contact with the outside world where the infection still spreads freely. Vaccine uptake rates are  $\alpha_1=\;\alpha_2=\;\alpha_3=\;0.9$.  People who have recovered from COVID-19 are assumed to retain natural immunity for 180 days \cite{Nature2020}, and the endemic circles in the case of  no vaccine are the result of this temporary natural immunity conferred by the disease.}
	\label{fig:doses_i}
	\vspace{-4mm} 
\end{figure*}

We simulated the disease dynamics in the population of the township, represented by the scale-free network as described above, for the following four cases: (i) no vaccination (ii) only one dose of vaccine administered (iii) at most two doses of vaccine administered (iv) at most three doses of vaccine administered. The variation of Susceptible, Infected (Exposed + Quarantined) and Vaccinated proportions of the population against time for each of these four cases are shown in \cref{fig:doses_i}.  It should be noted that in our model, people leave the Vaccinated ($V$) compartment when their immunity from vaccination wears off, so 'Vaccinated' should be taken to mean people who presently have immunity from vaccination, rather than people who have ever taken the vaccine. It could be noted from  \cref{fig:doses_i} that due to people losing natural immunity obtained from the disease over time, and due to people travelling outside of the town and some of them returning infected, the disease goes through endemic cycles and cannot be completely eradicated. However, the amplitude of the endemic cycle depends on how many vaccines are administered.  As shown in \cref{fig:no_dose_i}, when there is no vaccination, the epidemic peaks with about 90\% of the people infected, soon after the infection enters the town. Then the number of infected people drops sharply, yet after a while people lose their natural immunity and become susceptible again, and the infection peaks again. Yet, every infection peak has  a smaller amplitude compared to the previous one, due to more and more people retaining their natural immunity after being exposed to multiple infections. The infection peaks have a period of about 220 days, and eventually about 20\% of the community is infected at each peak.  These observations are consistent with the observed peaks of COVID-19 in communities which have limited exposure to the outside world and have not had any vaccination\cite{OWIDCoronavirus}. 

We also observe that if vaccinations are administered,  then the amplitude of the peaks decrease. Yet, we could observe from \cref{fig:one_dose_i}, when only one vaccine dose is administered, the peak amplitude is similar to the case where there is no vaccination. Evidently one vaccine hardly makes any difference in disease dynamics, due to the fact that immunity from a single vaccine is lost quickly. Bearing in mind that we model limited vaccination supply, it could be observed that the proportion of people in Vaccinated ($V$) compartment never increases to be more than 60\%. Even though the uptake rate is modelled as 90\%, the limited supplies mean that the first people who received vaccination have lost or about to lose the vaccination immunity by the time the last people obtain vaccination. Therefore, without follow up vaccination, the pathogen always finds plenty of susceptible people in the community. This matches with real-world observations that a single dose of vaccination hardly makes any positive difference in the long term \cite{OWIDCoronavirus}. We could also observe that the period of the endemic cycles remain at around $220$ days.

There is significant difference however, when two or three vaccines are administered, as \cref{fig:two_dose_i} and \cref{fig:three_dose_h3_i} show. In these cases it could be observed that the amplitude of the peak of the endemic cycles is significantly reduced, to about 10\%, and when three doses  are administered, the endemic cycles nearly disappear,  though the disease is not eradicated. What is significant to observe is that even though the third vaccine is administered 180 days (six months) after the second vaccine, the proportion of people retaining immunity (the `Vaccinated' compartment) remains significant long after that. In case of two vaccines being administered, about 30\% of the population retains immunity after 1080 days (nearly three years), while when three vaccines are administered, about half of the population retains immunity after three years. Thus, while the disease cannot be eradicated with three vaccines, they are sufficient to contain the disease for about three years, even when there is a level of contact with the outside world, where it is assumed that the disease still spreads freely.  It could be postulated therefore that if outside world is sufficiently vaccinated by this time, three vaccinations could be indeed enough to eradicate the disease completely from this township.

\begin{figure*}[h!]
	\centering

	\begin{subfigure}{0.28\textwidth}
		\includegraphics[width=\linewidth]{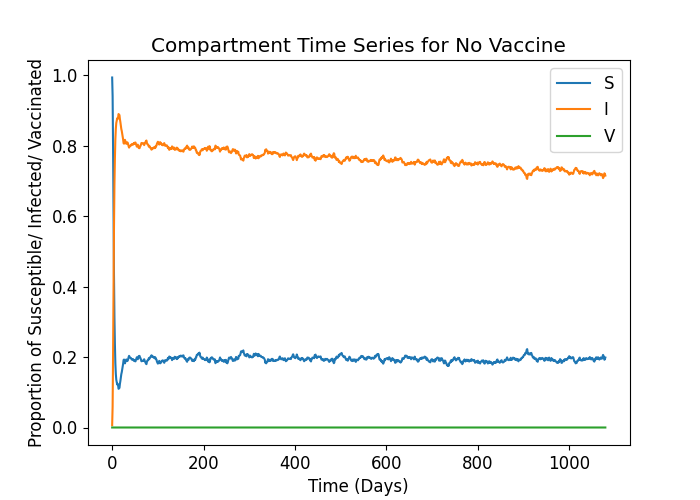}
		\caption{No vaccine provided}
		\label{fig:no_dose_immune_i}
	\end{subfigure}
	~
	\centering
	\begin{subfigure}{0.28\textwidth}
		\centering
		\includegraphics[width=\linewidth]{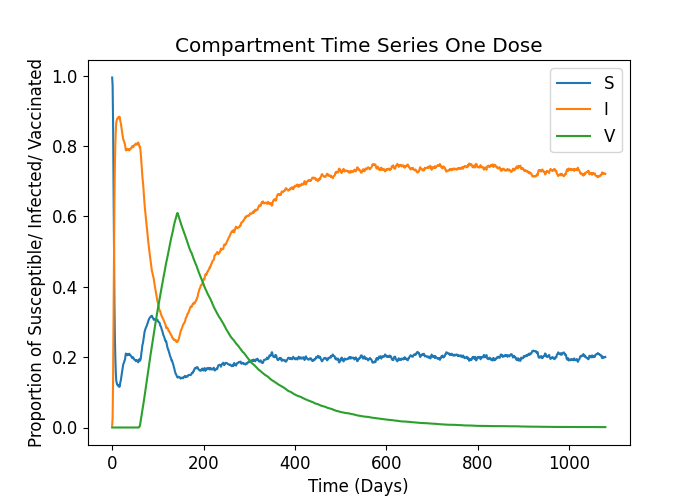}
		\caption{One dose}
		\label{fig:one_dose_immune_i}
	\end{subfigure}
	
	\centering
	\begin{subfigure}{0.28\textwidth}
		\centering
		\includegraphics[width=\linewidth]{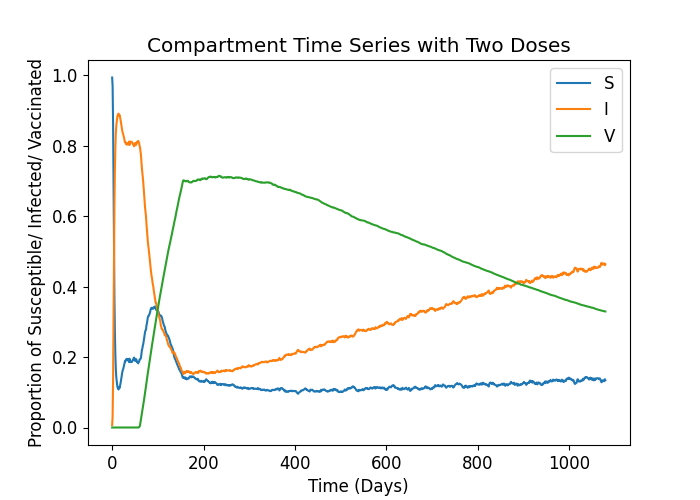}
		\caption{Two doses}
		\label{fig:two_dose_immune_i}
	\end{subfigure}
	~
	\centering
	\begin{subfigure}{0.28\textwidth}
		\centering
		\includegraphics[width=\linewidth]{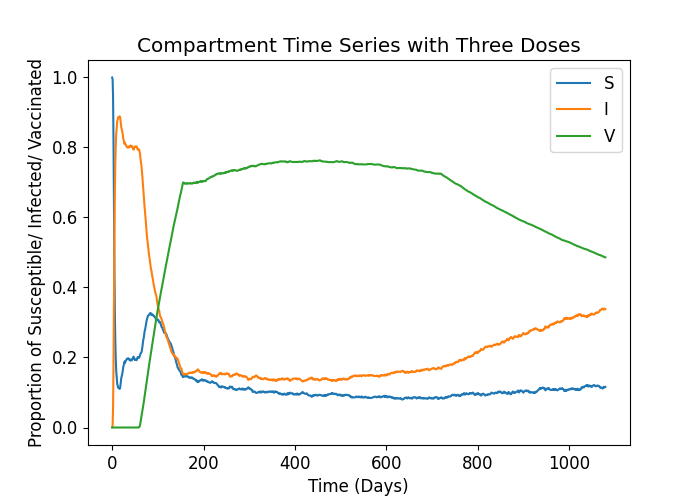}
		\caption{Three doses}
		\label{fig:three_dose_immune_i}
	\end{subfigure}
	\caption{Infectious disease dynamics of COVID-19 in the NSW township of 10,000 people when a) no vaccine b) one vaccine c) two vaccines d) three vaccines are administered. Second and third vaccines are administered 28 and 208 days after the first, respectively.   The proportion of susceptible ($S$), Infected ($I =  E + U$) and Vaccinated ($V$) people in the population are shown. Vaccine uptake rates are  $\alpha_1\;=\;\alpha_2\;=\;\alpha_3\;=\;0.9$.  In this case, the period of natural immunity conferred by contracting the disease varies such that a) people who have had no vaccine and no previous  COVID-19  infection retain natural  immunity for 140 days, people who have had one or more vaccines retain natural  immunity for 180 days,   people who have had  previous COVID-19 infection retain natural immunity for 180 days, and people who have had one or more vaccination and previous COVID-19 infection retain natural immunity for 200 days. Compare this with   \cref{fig:doses_i} where everyone who recovered from COVID-19 is assumed to have natural immunity for 180 days, regardless of previous history or vaccination status. }
	\label{fig:doses_immune_i}
		\vspace{-2mm} 
\end{figure*}

One limitation of these experiments has been that we assumed that everyone who had contracted COVID-19 will retain natural immunity (remain in `Recovered' ($R$) state)  for 180 days, after which they will return to the susceptible ($S$) state. That is, the `re-susceptibility rate' $\gamma$ is $\frac{1}{180}$. However, it has been shown that the re-susceptibility rate can depend on whether a person has previously had COVID-19, or whether that person has had vaccination. Therefore, we now  consider the case whereby there were multiple re-susceptibility rates. For this purpose, we conducted simulation experiments where the immunity period varies depending on the above factors. In particular, we assumed in these experiments that 1) people who have not had vaccination and have not had COVID-19 before have a natural immunity period of 140 days after COVID-19 infection: that is, the  `re-susceptibility rate'  $\gamma_0$ is $\frac{1}{140}$ 2) people who have not had vaccination but have had COVID-19 previously  have  a natural immunity period of  180 days after the  later COVID-19 infection: that is, the  `re-susceptibility rate'   $\gamma_c$ is $\frac{1}{180}$  3) people who have had one or more doses of vaccination  but have not had COVID-19 before have a natural immunity period of 180 days  after COVID-19 infection: that is, the  `re-susceptibility rate'  $\gamma_v$ is $\frac{1}{180}$  4) people who have had one or more doses of vaccination and have had COVID-19 previously  have  a natural immunity period of  200 days after the  later COVID-19 infection: that is, the  `re-susceptibility rate'  $\gamma_{vc}$ is $\frac{1}{200}$. Therefore, the immunity period now depends on a set of rules rather than being  equal for every member of the population.  Under these assumptions, we again considered disease dynamics for the four cases: a) no vaccination b) a maximum of one vaccine c) a maximum of two vaccines d) a maximum of three vaccines. The results are shown in  \cref{fig:doses_immune_i}.

We could observe from \cref{fig:doses_immune_i} that the variation of natural  immunity period depending on previous history of individuals significantly changes disease dynamics. The main observation is that endemic cycles are now absent, and in the case where there is no vaccination, after the initial infection peak the community recovers only very slowly, and the disease remains at epidemic level for a long time. When there is vaccination the infection levels drop after initial vaccination but increase back as the immunity obtained from vaccination reduces. Basically, regardless of the number of  vaccines, the number of people with immunity from vaccine decides the number of people with infection at any one point in time. The susceptible population remains fairly low, around 20\%,  after the initial infection peak. The main observation therefore is that if there is heterogeneity in natural immunity, the pathogen is able to find people who have low natural immunity and spread quickly in the community through them, whereby a more homogenous  distribution of natural immunity seems to prevent infection spread even when the number of susceptible people increase, but this happens only for a certain period after which the pathogen is able to spread again, resulting in endemic cycles as shown in  \cref{fig:doses_i}. Therefore we postulate that a more heterogenous distribution of natural immunity assists quick infection spread, whereas a more homogenous distribution of natural immunity has a delaying effect,   resulting in endemic cycles.

\subsection{Cross-correlation between hub-infection and overall infection }

\begin{figure*}[h!]
	\centering
	\begin{subfigure}{0.28\textwidth}
		\centering
		\includegraphics[width=0.9\linewidth]{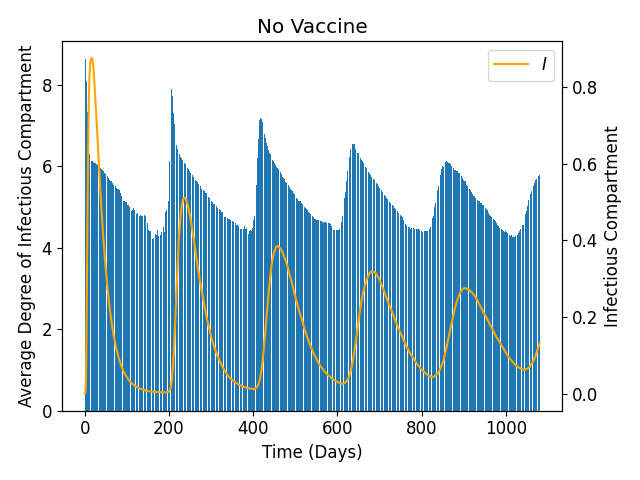}
		\caption{No vaccine provided}
		\label{fig:no_dose_ik}
	\end{subfigure}
	~
	\begin{subfigure}{0.28\textwidth}
		\centering
		\includegraphics[width=0.9\linewidth]{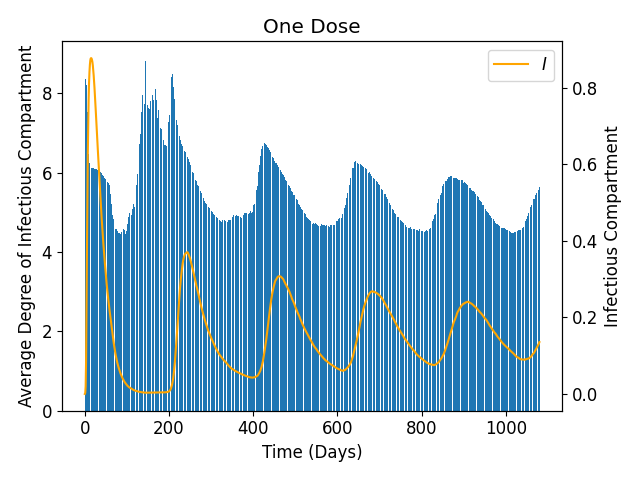}
		\caption{One dose}
		\label{fig:one_dose_ik}
	\end{subfigure}
	
	\begin{subfigure}{0.28\textwidth}
		\centering
		\includegraphics[width=0.9\linewidth]{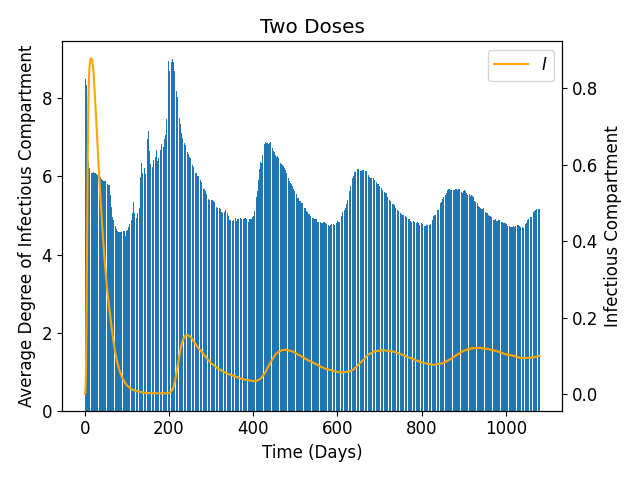}
		\caption{Two doses}
		\label{fig:two_dose_ik}
	\end{subfigure}
	~
	\begin{subfigure}{0.28\textwidth}
		\centering
		\includegraphics[width=0.9\linewidth]{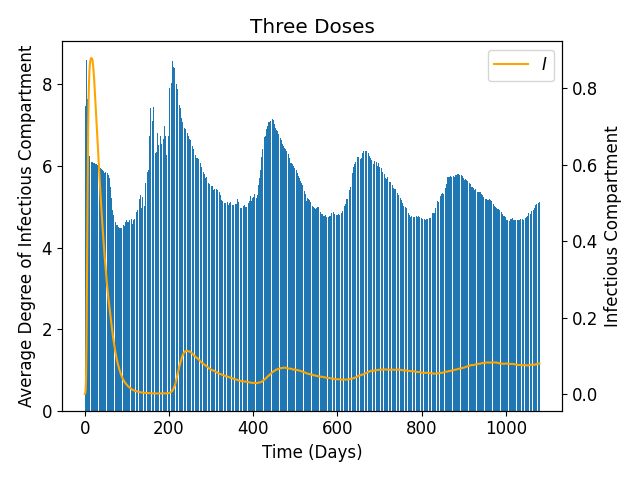}
		\caption{Three doses}
		\label{fig:three_dose_h3_ik}
	\end{subfigure}
	\caption{$\left\langle k\right\rangle_I$ and $I$ compared over time for  a) no vaccine b) one vaccine c) two vaccines d) three vaccines. Blue bars represent the average degree of infected  population $\left\langle k\right\rangle_I$, and the orange line represents the proportion of infected population $I$.  There is clear cross-correlation between the two time series,  with a time lag. The results correspond to simulations shown in \cref{fig:doses_i}. }
	\label{fig:ik}
	\vspace{-5mm} 
\end{figure*}

Here we analysed whether there is any cross correlation between the hubs of the contact network being infected, and the overall level of infection in the contact network of the  township.  The  motivation for this is to see whether high levels of infection of the hubs can be used to predict an overall rise  of infection levels in the community, and thereby to understand if targeted vaccination of the hubs could help prevent infection peaks. Rather than defining `hubs' as nodes which have degrees higher than a certain threshold, we simply looked at the average degree of the infected nodes, $\left\langle k\right\rangle_I$, and its cross correlation with the proportion of infected nodes $I$, over time. The reason for this is to avoid having to arbitrarily  decide on a degree threshold for `hubs', in a network with a fairly heterogenous degree distribution.  These two quantities are plotted in \cref{fig:ik}, for the case of a) no vaccine b) one vaccine c) two vaccines d) three vaccines as before.

It can be seen from the figures that there appear to be cross-correlation between these two quantities: $\left\langle k\right\rangle_I$ and $I$. That is, when infection of nodes with higher degrees peak, an overall infection peak soon follows. It could also be observed that the lag time between these two sets of peaks is longer when there is less vaccination (zero or one dose), and shorter when there is more vaccination (two or three doses). Of course, when there are more doses administered the infection peaks are less pronounced, and consequently the cross-correlation is  less pronounced. We further quantify this cross-correlation by fitting the time series of $\left\langle k\right\rangle_I$ and $I$ into the Vector autoregression (VAR) model. For this, a univariate autoregression of $I_t$ is constructed  and then augmented by including the lagged values of  ${\langle k \rangle}_{I,t}$ as:

\vspace{-4mm} 

%
%\[I_t\;=\;A_1I_{t\;-\;1}\;+\;A_2I_{t\;-\;2}\;+\;\cdots\;+\;A_pI_{t\;-\;p}\;+\;\cdots\;+\;A_0I_0\;+\;c\]
%
%and then augmented by including the lagged values of  ${\langle k \rangle}_{I,t}$

\begin{align}
I_t\;&=\;A_1I_{t\;-\;1}\;+\;A_2I_{t\;-\;2}\;+\;\cdots\;+\;A_pI_{t\;-\;p}\;+\;\cdots\;+\;A_0I_0\;  + \nonumber  \\
\;&+ B_p {\langle k \rangle}_{I , t-p}+  \;\cdots\; + B_q {\langle k \rangle}_{I,t-q} +\;c\ \nonumber
\end{align}

%\begin{align}
%	\left\langle k\right\rangle_{I,\,t}\;&=\;A_1\left\langle k\right\rangle_{I,\,t\;-\;1} \;+\;A_2\left\langle k\right\rangle_{I,\,t\;-\;2} \;+\;\cdots\;+\;A_p\left\langle k\right\rangle_{I,\,t\;-\;p} \;+\;\cdots \nonumber\\
%	\;&+\;A_0\left\langle k\right\rangle_{I,\,0}\nonumber
%\end{align}

The model fits the data based on the maximum lag $p$. We then conducted the Granger causality test \cite{Diebold1998},  to test whether one time series (e.g. $\left\langle k\right\rangle_{I,t}$) can predict the other (i.e. $I_t$). The \cref{table:GrangerTest} shows the results of the Granger causality test. Noting that scenarios  with a $p$-value less than the $0.05$ threshold imply a statistically significant causality (where the null hypothesis of no causality is rejected), we could note that in the case of no vaccines being administered, the timeseries of $\left\langle k\right\rangle_I$ predicts  $I$. That is, in this case, if the average degree of  infected people increases, it is an indication that the infection numbers are going to peak. This is born out also by  \cref{fig:ik}  which shows obvious cross correlation when no vaccines are administered. When one, two, or three vaccines are administered,  again the $p$-value is very close to zero, and it decreases further as the number of vaccinations increase. Therefore when vaccinations have been administered also, the average degree of infectious people can predict the infection numbers, and again this is  born out  by  \cref{fig:ik}. Using the lowest Akaike information criterion (AIC)\cite{Diebold1998}, we also calculated the lag time, which is also shown in \cref{table:GrangerTest}.  It could be seen that the higher the doses, the lower the lag time: that is, the peaks of the infection  follow  $\left\langle k\right\rangle_I$,  \cref{fig:ik}, with smaller and smaller lags.  When a maximum of one dose is administered, the lag is 31 days, while when a maximum of two doses are administered, the lag is 11 days, and when a maximum of three doses are administered, the lag is 9 days.  These results show that if hubs in the contact network get infected, this is likely to result in the overall levels of infection peaking, regardless of vaccination saturation, and this can even be used as an early warning system, if we  can identify the people in the community who are the hubs in the contact networks (have the most contacts).

\begin{table}[h!]
	\centering
	\begin{tabularx}{0.45\textwidth}{ | X | c | c | }
		\hline
		\emph{Case} & \emph{Lowest AIC (Lag)} & $p$ \\ 
		\hline
		Three doses & $-9$ & $\num{5.020e-7}$ \\
		\hline
		Two doses & $-11$ & $\num{8.18e-6}$\\
		\hline
		One dose & $-31$ & $\num{2.32e-5}$\\
		\hline
		No vaccines & $-250$ & $0.012$\\
		\hline
	\end{tabularx}
	\caption{Results from Granger Tests, comparing times series $\left\langle k\right\rangle_I$ and $I$, using the  lowest Akaike information criterion (AIC).  The negative values of lag times indicate that the  time series $\left\langle k\right\rangle_I$  leads  time series  $I$.  A value of $p \le 0.05$ indicates statistical significance.  Therefore all four cases are statistically significant.}
	\label{table:GrangerTest}
	\vspace{-2mm} 
\end{table}

\begin{figure}[h!]
	\vspace{-3mm} 
		\centering
		\includegraphics[width=0.8\linewidth]{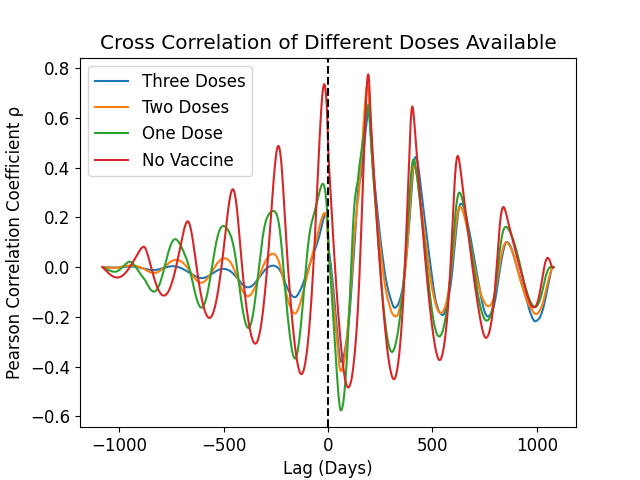}
	\caption{Cross-correlation coefficient   $\rho$ against  lag time for the time series $\left\langle k\right\rangle_I$ and $I$ for the four distinct cases: a) zero  b) one c) two d) three vaccines. In all cases, there is strong cross-correlation which peaks for a lag time between 25 and 12 days. (A negative lag time indicates $\left\langle k\right\rangle_I$  leads $I$.   Since we are interested in predicting $I$, not  $\left\langle k\right\rangle_I$,  we only consider negative values for the lag-time.)}
	\label{fig:ccf}
	\vspace{-5mm} 
\end{figure}

We also  directly computed the cross-correlation $\rho$   between the time series $\left\langle k\right\rangle_I$ and $I$. We calculated this cross correlation coefficient  $\rho$ for a range of lag times, and also for the four cases as before: a) no vaccination b) one vaccine c) two vaccines 3) three vaccines. The results are shown in \cref{fig:ccf}, from which we can identify what are the lag times for which the maximum cross correlation occurs. Note here that even though the figure shows both positive lag times and negative lag times, we are only interested in the negative lag times, since we wish to see if  $\left\langle k\right\rangle_I$ leads (can predict) $I$, and not the other way around. We could observe that for the no vaccine case, the maximum cross-correlation between the time series  $\left\langle k\right\rangle_I$ and $I$ is 0.76, and that occurs when the lag time is -25 days ($\left\langle k\right\rangle_I$ leads $I$ by 25 days). Similarly, for the (maximum of) one vaccine case,  the maximum cross-correlation between the time series  $\left\langle k\right\rangle_I$ and $I$ is 0.36, and that occurs when the lag time is  -18 days. For the (maximum of)  two vaccines case, the maximum cross-correlation between the time series  $\left\langle k\right\rangle_I$ and $I$ is 0.22, and that occurs when the lag time is  -16 days. Finally, for the (maximum of)  three vaccines case,  the maximum cross-correlation between the time series  $\left\langle k\right\rangle_I$ and $I$ is 0.21, and that occurs when the lag time is  -12 days. These results can be interpreted in a way similar to the Granger causality test results, and again we see strong cross-correlation in all cases.  Therefore both the Granger test and direct cross-correlation calculation give qualitatively very similar results and  show that infection of the higher-degreed nodes (hubs) can predict overall infection peaks in this community, and as more vaccines are administered, the lead  time of the predictor reduces.

\section{Conclusion}

In this work, we considered a NSW township of 10,000 people, which has limited contact with the rest of Australia, and modelled COVID-19 disease dynamics and vaccination uptake in this township using an agent-based contact network model. We simulated upto three doses of vaccination, with limited vaccination supply at the beginning of the vaccination program. For each dose, we assumed 90\% vaccination uptake compared to the previous dose, and we also assumed that natural immunity obtained from exposure to SARS-CoV-2  will  last for a time. Using a parameterised compartmental model, we analysed the effects of vaccination on disease dynamics, with the goal of understanding whether the disease can be eradicated or contained in this township with limited outside contact with a maximum of three vaccines. Especially, we compared the scenario where the waning rate of natural immunity conferred by the disease is constant, to the scenario where this waning rate depends on whether the individual concerned had COVID-19  and/or  vaccination before. We assumed constant waning rates for immunity conferred by the vaccines. We also studied the relationship between infection levels in higher-degreed nodes (more connected individuals)  and overall infection levels. We used the efficacy characteristics of Pfizer Comirnaty (BNT162b2)  mRNA  vaccine in our study.

The results  indicate that in such a township, COVID-19 can be contained, but not eradicated, by up to three  vaccinations.  The disease remains endemic, however the disease dynamics depend  on the nature of natural immunity conferred by the disease. If the natural immunity conferred by COVID-19 is homogenous  and does not depend on the disease and vaccination history of individuals, the township faces endemic circles of the disease, though the amplitude of these circles  lessen if more vaccines are administered. If the  natural immunity conferred by the disease is heterogenous and depends on the disease and vaccination history of individuals, then the disease remains at epidemic levels for a long time and can only be contained slowly. In this case, we postulate that the pathogen is able to find individuals with less immunity and propagate through them, affecting other individuals as their vaccination-conferred  immunity wanes, and more vaccines are needed to effectively contain the disease.

We also showed that there is a causal relationship between the average degree of infected people and the number of (or proportion of) infected people in the town. If this average degree increases, it predicts a peak in overall infection levels. We calculated cross-correlation coefficients and employed Granger causality test to demonstrate this. The results of either method are qualitatively similar. If more vaccines are administered, the lag  between the average degree of infected nodes (people), the predictor, and the overall infection proportion also decreases,  so that the the average degree of the infected people is perhaps more useful as a short term predictor. Essentially, these experiments indicated that if the more connected people in the town are highly infected compared to the rest of the populace, this indicates an oncoming peak of the infection levels overall. 

While the effects of vaccination on disease dynamics in the context of COVID-19 has been studied extensively, this study has focused on a small township which has limited  outside contact. While that would be relevant to small townships all over the world, we have made the simulation realistic particularly for a hypothetical NSW township in Australia, in terms of vaccine uptake, vaccine type, and  disease  parameters. Overall, the main contribution of this study is to computationally model COVID-19 disease dynamics,  particularly during a sustained vaccination program, in a typical Australian country town, and demonstrate how the epidemic would respond to such a program and what lessons can be learnt from it. Nevertheless, the model could be easily  adapted to small remote towns elsewhere by calibrating these parameters according to the local demographics. Therefore  the study provides important insights into the relationship between vaccination uptake and disease dynamics for  such a remote town, regardless of the country.

\printbibliography

\end{document}